# On the Foundational Equations of the Classical Theory of Electrodynamics

Masud Mansuripur

College of Optical Sciences, The University of Arizona, Tucson, Arizona 85721

**Abstract**. A close examination of the Maxwell-Lorentz theory of electrodynamics reveals that polarization and magnetization of material media need *not* be treated as local averages over small volumes – volumes that nevertheless contain a large number of electric and/or magnetic dipoles. Indeed, Maxwell's macroscopic equations are exact and self-consistent mathematical relations between electromagnetic fields and their sources, which consist of free charge, free current, polarization, and magnetization. When necessary, the discrete nature of the constituents of matter and the granularity of material media can be handled with the aid of special functions, such as Dirac's delta-function. The energy of the electromagnetic field and the exchange of this energy with material media are treated with a single postulate that establishes the Poynting vector $\boldsymbol{S}=\boldsymbol{E}\times\boldsymbol{H}$ as the rate of flow of electromagnetic energy under all circumstances. Similarly, the linear and angular momentum densities of the fields are simple functions of the Poynting vector that can be unambiguously evaluated at all points in space and time, irrespective of the type of material media, if any, that might reside at various locations. Finally, we examine the Einstein-Laub force- and torque-density equations, and point out the consistency of these equations with the preceding postulates, with the conservation laws, and with the special theory of relativity. The set of postulates thus obtained constitutes a foundation for the classical theory of electrodynamics.

**1. Introduction**. I would like to invite you, the reader, to temporarily abandon what you have learned in your college courses about electrodynamics. Sometimes it is hard to reconcile a new outlook with what one believes to be the absolute truth. I will try to cover the foundations of classical electrodynamics in just a few pages. Even when you find some of the statements on these pages to contradict your most cherished notions of the Maxwell-Lorentz theory of electromagnetism, I hope you will be kind enough to bear with me until the end.

The International System of Units (abbreviated SI from the French *Système international d'unités*) is used throughout this article. Electromagnetic (EM) phenomena occur in space and time, and we assume that the space-time is flat, with each point $\boldsymbol{r}=(x,y,z)$ in space having its own clock that marks the passage of time $t$ at a fixed rate. An inertial observer (i.e., one traveling at a constant velocity relative to distant stars[†]) observes the EM phenomena as they occur in this flat space-time. The clocks are all synchronized for this observer, and each point in space-time is specified by its $(\boldsymbol{r},t)$ coordinates. The rules of special relativity will be assumed to apply, so that all phenomena in a given inertial frame can be equivalently described in any other inertial frame using the Lorentz transformation of space-time coordinates as well as those of energy, momentum, charge, current, EM fields, and so on.

The electromagnetic vacuum (or free space) is defined by two parameters, $\varepsilon_0 \approx 8.854\times10^{-12}$ farad/meter and $\mu_0 = 4\pi\times10^{-7}$ henry/meter, known respectively as the permittivity and permeability of free space. These two parameters remain the same for all inertial observers. The speed of light in vacuum is given by $c = 1/\sqrt{\mu_o\varepsilon_o} \approx 3\times10^8$ m/s, while the impedance of free space is $Z_o = \sqrt{\mu_o/\varepsilon_o} \approx 377\,\Omega$. In classical electrodynamics, we have four sources for the EM fields. These sources, which are generally represented by continuous and differentiable functions of space-time are: (i) free charge density $\rho_{\text{free}}(\boldsymbol{r},t)$ [coulomb/m$^3$], (ii) free current density $\boldsymbol{J}_{\text{free}}(\boldsymbol{r},t)$ [ampere/m$^2$], (iii) polarization $\boldsymbol{P}(\boldsymbol{r},t)$ [coulomb/m$^2$], and (iv) magnetization $\boldsymbol{M}(\boldsymbol{r},t)$ [weber/m$^2$]. Note that we are already departing from conventional wisdom by assuming $\boldsymbol{P}(\boldsymbol{r},t)$ and $\boldsymbol{M}(\boldsymbol{r},t)$ to be well-defined functions that are specified for all points $\boldsymbol{r}$ in space and instants $t$ in time. In other words, we are *not* subscribing to the notion that $\boldsymbol{P}$ is some sort of local average of electric dipole moments over a small volume of space, or, likewise, that $\boldsymbol{M}$ is a local average of magnetic dipole moments [1]. In our approach, the granularity of matter can be handled with the aid of special functions such as Dirac's delta-function $\delta(\boldsymbol{r})$. A point-charge $q$ moving at a constant velocity $\boldsymbol{V}$, for example, is described by $\rho_{\text{free}}(\boldsymbol{r},t) = q\delta(\boldsymbol{r}-\boldsymbol{V}t)$ and $\boldsymbol{J}_{\text{free}}(\boldsymbol{r},t) = q\boldsymbol{V}\delta(\boldsymbol{r}-\boldsymbol{V}t)$. Similarly, a magnetic point-dipole $\boldsymbol{m}_o$ located at $\boldsymbol{r} = \boldsymbol{r}_o$ and oscillating at a fixed frequency $\omega_o$ is represented by $\boldsymbol{M}(\boldsymbol{r},t) = \boldsymbol{m}_o\delta(\boldsymbol{r}-\boldsymbol{r}_o)\sin(\omega_o t)$.

[†] Strictly speaking, this notion of an inertial observer is no longer valid, because distant stars and galaxies are not stationary. The modern definition describes inertial frames as frames in which the laws of physics are simple.

In addition to material media, which, in classical electrodynamics, are represented by the four sources $\rho_{\text{free}}$, $\boldsymbol{J}_{\text{free}}$, $\boldsymbol{P}$ and $\boldsymbol{M}$, there exist the EM fields $\boldsymbol{E}(\boldsymbol{r},t)$ [volt/m] and $\boldsymbol{H}(\boldsymbol{r},t)$ [ampere/m]. These vector functions of the space-time differ substantially from the sources in that they are *not* matter in the ordinary sense. The fields are produced by the sources, of course; they contain energy and momentum, and are capable of exchanging their energy and momentum with material media. Two other fields that appear in Maxwell's equations are the displacement $\boldsymbol{D}(\boldsymbol{r},t) = \varepsilon_o\boldsymbol{E}(\boldsymbol{r},t) + \boldsymbol{P}(\boldsymbol{r},t)$ [coulomb/m$^2$] and the magnetic induction $\boldsymbol{B}(\boldsymbol{r},t) = \mu_o\boldsymbol{H}(\boldsymbol{r},t) + \boldsymbol{M}(\boldsymbol{r},t)$ [weber/m$^2$]. In the special theory of relativity, $\boldsymbol{E}$ and $\boldsymbol{B}$ form a second-rank tensor that can be transformed between different inertial frames using the Lorentz transformation rules. Similarly, $\boldsymbol{D}$ and $\boldsymbol{H}$ form a second rank tensor, as do $\boldsymbol{P}$ and $\boldsymbol{M}$. It is thus natural to consider $\boldsymbol{E}$ and $\boldsymbol{B}$ as pure EM fields, with the composite fields $\boldsymbol{D}$ and $\boldsymbol{H}$ constructed as mixtures of pure fields and source densities $\boldsymbol{P}$ and $\boldsymbol{M}$. This division of the four fields $(\boldsymbol{E},\boldsymbol{D},\boldsymbol{H},\boldsymbol{B})$ into pure and composite is an aesthetic matter with no practical consequences.

**2. Constitutive relations**. A source, such as a point-charge or a magnetic particle, can be driven by an external agent or by the EM fields produced by existing sources within a given system. An electron affixed to a proverbial ant and moving about at the ant's whim is an example of a source driven by an external agent, whereas two protons flying apart are responding to the EM fields that their own motion has created. In the electrodynamics of Maxwell and Lorentz, the spatio-temporal distribution of the sources may be predetermined, as when an electrically-charged plastic rod is set in oscillatory motion by a mechanical device. Alternatively, the source and the radiation may be interdependent, as when a pulse of light is trapped between two parallel mirrors, where the $E$-field of the light pulse excites the conduction electrons on the mirror surfaces, while the oscillating electrons sustain the $E$- and $H$-fields that comprise the light pulse.

As a general rule, Maxwell's equations are agnostic about the mechanism(s) by which the sources are set in motion. These equations relate the EM fields $\boldsymbol{E}$ and $\boldsymbol{H}$ and their various derivatives to the local sources $\rho_{\text{free}}$, $\boldsymbol{J}_{\text{free}}$, $\boldsymbol{P}$, $\boldsymbol{M}$ and their derivatives. When the sources are predetermined, one simply solves Maxwell's equations to obtain the resulting EM fields. It may happen, however, that the force and torque exerted by the fields on the sources dictate the motion of the material media in the presence of the fields, in which case a self-consistent solution of Maxwell's equations along with the relevant equations of motion of the media become necessary. A third possibility is that the response of the material media to EM fields is more complicated, involving, for example, quantum-mechanical interactions among the various constituents of matter. This happens, for instance, when electric (magnetic) dipoles oscillate in response to oscillating local electric (magnetic) fields, or when conduction electrons follow the variations of the local $E$-field in accordance with Ohm's law. The response of the material media to EM fields is often specified via constitutive relations, e.g., Ohm's law $\boldsymbol{J}_{\text{free}}(\boldsymbol{r},\omega)=\sigma(\boldsymbol{r},\omega)\boldsymbol{E}(\boldsymbol{r},\omega)$, where $\sigma(\boldsymbol{r},\omega)$ is the electrical conductivity of a medium at point $\boldsymbol{r}$ in space in response to a local $E$-field that oscillates at frequency $\omega$. Other examples of constitutive relations for linear materials are $\boldsymbol{P}(\boldsymbol{r},\omega)=\varepsilon_{\text{o}}\chi_{\text{e}}(\boldsymbol{r},\omega)\boldsymbol{E}(\boldsymbol{r},\omega)$ and $\boldsymbol{M}(\boldsymbol{r},\omega)=\mu_{\text{o}}\chi_{\text{m}}(\boldsymbol{r},\omega)\boldsymbol{H}(\boldsymbol{r},\omega)$, with $\chi_{\text{e}}$ and $\chi_{\text{m}}$ being, respectively, the relative electric and magnetic susceptibility of the material [1-4].

Bear in mind that the materials may be inhomogeneous, anisotropic, nonlinear, dispersive, absorptive, hysteretic, etc., in which cases the constitutive relations may become extremely complicated. There exist theoretical models, such as the Lorentz oscillator model [4,5], that relate the response of various sources to the EM fields, but also empirical or heuristic formulas based on experimental determination of the constitutive relations that are often used in practice. As important as the constitutive relations are in enabling the solution of Maxwell's equations in practical situations, they are *not* fundamental to the theory of electrodynamics, as they are usually rooted in other branches of physics (e.g., solid state, statistical, quantum). In the remainder of this paper, we shall not discuss the constitutive relations, nor, for that matter, other mechanisms that might affect the spatio-temporal behavior of the sources in the presence of EM fields; rather, we shall assume that the relevant constitutive relations and/or equations of motion have already been accounted for, and that the functions $\rho_{\text{free}}(\boldsymbol{r},t)$, $\boldsymbol{J}_{\text{free}}(\boldsymbol{r},t)$, $\boldsymbol{P}(\boldsymbol{r},t)$ and $\boldsymbol{M}(\boldsymbol{r},t)$ appearing in various equations reflect the behavior of the material media in the presence of EM fields, which fields are generally produced (in a closed system) by the same sources. This is tantamount to running an experiment during which all the sources are continually monitored, then using the measured source distributions throughout space and time in order to compute the EM fields $\boldsymbol{E}(\boldsymbol{r},t)$ and $\boldsymbol{H}(\boldsymbol{r},t)$ as they existed during the experiment.

**3. Maxwell's macroscopic equations**. We are now ready to discuss Maxwell's equations. These equations, which relate the fields to their sources, are commonly known as the "macroscopic" equations, distinct from the "microscopic" equations, which are considered by almost all physicists to be the more fundamental equations. Here, however, we shall treat the macroscopic equations of Maxwell as fundamental, in the sense that they incorporate polarization $\boldsymbol{P}$ and magnetization $\boldsymbol{M}$ as distinct sources on a par with charge and current densities $\rho_{\text{free}}$ and $\boldsymbol{J}_{\text{free}}$. In their differential form, Maxwell's macroscopic equations are written [1-4]:

$$\boldsymbol{\nabla}\cdot\boldsymbol{D}(\boldsymbol{r},t)=\rho_{\text{free}}(\boldsymbol{r},t), \tag{1}$$

$$\boldsymbol{\nabla}\times\boldsymbol{H}(\boldsymbol{r},t)=\boldsymbol{J}_{\text{free}}(\boldsymbol{r},t)+\partial\boldsymbol{D}(\boldsymbol{r},t)/\partial t, \tag{2}$$

$$\boldsymbol{\nabla}\times\boldsymbol{E}(\boldsymbol{r},t)=-\partial\boldsymbol{B}(\boldsymbol{r},t)/\partial t, \tag{3}$$

$$\boldsymbol{\nabla}\cdot\boldsymbol{B}(\boldsymbol{r},t)=0. \tag{4}$$

An immediate consequence of these equations is the charge-current continuity equation, which is obtained by applying the divergence operator to both sides of Eq.(2), then replacing for $\boldsymbol{\nabla}\cdot\boldsymbol{D}$ from Eq.(1) to arrive at

$$\boldsymbol{\nabla}\cdot\boldsymbol{J}_{\text{free}}+\partial\rho_{\text{free}}/\partial t=0. \tag{5}$$

Clearly $\rho_{\text{free}}$ and $\boldsymbol{J}_{\text{free}}$ are interdependent; in other words, one cannot specify the spatio-temporal distributions of these two sources independently. Another constraint is that $(\boldsymbol{J}_{\text{free}}, c\rho_{\text{free}})$ behaves as a 4-vector, which fully specifies the charge and current distributions in all inertial frames once they are given in one such frame [1].

Maxwell's macroscopic equations are reduced to his microscopic equations if we *define* the bound charge and current densities $\rho_{\text{bound}}(\boldsymbol{r},t) = -\boldsymbol{\nabla}\cdot\boldsymbol{P}$ and $\boldsymbol{J}_{\text{bound}}(\boldsymbol{r},t) = (\partial\boldsymbol{P}/\partial t) + \mu_{\text{o}}^{-1}\boldsymbol{\nabla}\times\boldsymbol{M}$, and use them to eliminate $\boldsymbol{D}$ and $\boldsymbol{H}$ from Eqs. (1) and (2). We will find

$$\varepsilon_{\text{o}}\boldsymbol{\nabla}\cdot\boldsymbol{E}(\boldsymbol{r},t) = \rho_{\text{total}}(\boldsymbol{r},t), \tag{6}$$

$$\boldsymbol{\nabla}\times\boldsymbol{B}(\boldsymbol{r},t) = \mu_{\text{o}}\boldsymbol{J}_{\text{total}}(\boldsymbol{r},t) + \mu_{\text{o}}\varepsilon_{\text{o}}\partial\boldsymbol{E}(\boldsymbol{r},t)/\partial t. \tag{7}$$

In the above equations, $\rho_{\text{total}} = \rho_{\text{free}} + \rho_{\text{bound}}$ and $\boldsymbol{J}_{\text{total}} = \boldsymbol{J}_{\text{free}} + \boldsymbol{J}_{\text{bound}}$. (Equations (3) and (4) are shared between the macroscopic and microscopic forms.) It is easy to verify that the continuity equation, Eq. (5), is also satisfied by $\rho_{\text{bound}}$ and $\boldsymbol{J}_{\text{bound}}$, simply because of the way these bound entities have been defined. The continuity equation thus imposes no constraints on $\boldsymbol{P}(\boldsymbol{r},t)$ and $\boldsymbol{M}(\boldsymbol{r},t)$, in the sense that, mathematically speaking, these functions can have arbitrary values at various points in space-time. However, as pointed out earlier, $\boldsymbol{P}$ and $\boldsymbol{M}$ form a second-rank tensor that obeys the Lorentz transformation rules between inertial frames. Thus, once $\boldsymbol{P}(\boldsymbol{r},t)$ and $\boldsymbol{M}(\boldsymbol{r},t)$ are specified in one inertial frame, they will be known in all such frames. The Lorentz transformation rules also ensure that $(\boldsymbol{J}_{\text{bound}}, c\rho_{\text{bound}})$ acts as a 4-vector. It is for these reasons that Maxwell's "microscopic" equations are generally considered to be fundamental, with $\rho_{\text{total}}$ and $\boldsymbol{J}_{\text{total}}$ being the sole sources of the $\boldsymbol{E}$ and $\boldsymbol{B}$ fields. We, however, would like to retain $\boldsymbol{P}$ and $\boldsymbol{M}$ as distinct and separate sources, not because we disagree with the above treatment, but because, as will be seen shortly, the exchange of energy and momentum between fields and sources shows that bound charges and currents often behave differently than their free counterparts.

Let us also mention in passing that an alternative treatment of $\boldsymbol{P}$ and $\boldsymbol{M}$ in Maxwell's macroscopic equations involves the introduction of bound *magnetic* charge and current densities $\rho_{\text{bound}}^{(\text{m})} = -\boldsymbol{\nabla}\cdot\boldsymbol{M}$ and $\boldsymbol{J}_{\text{bound}}^{(\text{m})} = (\partial\boldsymbol{M}/\partial t) - \varepsilon_{\text{o}}^{-1}\boldsymbol{\nabla}\times\boldsymbol{P}$. In this way, one keeps Eqs. (1) and (2) as they are, but eliminates $\boldsymbol{E}$ and $\boldsymbol{B}$ from Eqs. (3) and (4), writing them in the following equivalent form:

$$\boldsymbol{\nabla}\times\boldsymbol{D}(\boldsymbol{r},t) = -\varepsilon_{\text{o}}\boldsymbol{J}_{\text{bound}}^{(\text{m})} - \mu_{\text{o}}\varepsilon_{\text{o}}\partial\boldsymbol{H}(\boldsymbol{r},t)/\partial t, \tag{8}$$

$$\mu_{\text{o}}\boldsymbol{\nabla}\cdot\boldsymbol{H}(\boldsymbol{r},t) = \rho_{\text{bound}}^{(\text{m})}(\boldsymbol{r},t). \tag{9}$$

As before, the magnetic bound charge and current densities satisfy the continuity equation, and $(\boldsymbol{J}_{\text{bound}}^{(\text{m})}, c\rho_{\text{bound}}^{(\text{m})})$ acts as a 4-vector [6]. There is absolutely no difference between treatments in which $\boldsymbol{P}$ and $\boldsymbol{M}$ are replaced by their equivalent bound *electric* charge and current densities, and those in which equivalent bound *magnetic* charge and current densities are employed. Some problems may be easier to solve using one method rather than the other, but the final results are always going to be the same.

**4. Electromagnetic energy and the Poynting vector**. There is nothing in Maxwell's equations per se that would hint at the energy associated with the EM fields and their sources. One needs the *postulate* of J. H. Poynting to begin to analyze the flow of energy in EM systems and the exchange of energy between the fields and the material media. According to Poynting's postulate, the rate of flow of EM energy per unit area per unit time is given by [1-3]:

$$\boldsymbol{S}(\boldsymbol{r},t)=\boldsymbol{E}(\boldsymbol{r},t)\times\boldsymbol{H}(\boldsymbol{r},t). \tag{10}$$

The above identity holds under all circumstances, whether the fields are in vacuum or inside material media, irrespective of the nature of the fields or the media. In other words, the fields may be static or dynamic, propagating or evanescent, and the media may be transparent or absorptive, linear or nonlinear, isotropic or birefringent, active or passive, homogeneous or inhomogeneous, etc. Poynting's famous theorem may subsequently be derived by dot-multiplying Eq. (2) with $\boldsymbol{E}(\boldsymbol{r},t)$ and Eq. (3) with $\boldsymbol{H}(\boldsymbol{r},t)$, then subtracting the two equations from each other [1-5]. One will find

$$\boldsymbol{\nabla}\cdot\boldsymbol{S}+\frac{\partial}{\partial t}(\tfrac{1}{2}\varepsilon_{\text{o}}\boldsymbol{E}\cdot\boldsymbol{E}+\tfrac{1}{2}\mu_{\text{o}}\boldsymbol{H}\cdot\boldsymbol{H})+\boldsymbol{E}\cdot\boldsymbol{J}_{\text{free}}+\boldsymbol{E}\cdot\frac{\partial\boldsymbol{P}}{\partial t}+\boldsymbol{H}\cdot\frac{\partial\boldsymbol{M}}{\partial t}=0. \tag{11}$$

Equation (11) is the continuity equation for EM energy. The first term in this equation, $\boldsymbol{\nabla}\cdot\boldsymbol{S}(\boldsymbol{r},t)$, is the rate of flow of EM energy out of a small volume $\delta v$ surrounding the point $\boldsymbol{r}$, evaluated at time $t$ and normalized by $\delta v$. The next term in Eq. (11) gives the time-rate-of-change of the EM energy density, $½\varepsilon_{\text{o}}|\boldsymbol{E}|^2+½\mu_{\text{o}}|\boldsymbol{H}|^2$, at $\boldsymbol{r}$ and $t$. Energy exchange between the EM fields and the free current density at $(\boldsymbol{r},t)$ occurs at a rate given by $\boldsymbol{E}\cdot\boldsymbol{J}_{\text{free}}$ per unit time per unit volume. If this dot-product happens to be positive, the energy flows from the fields into the material medium (i.e., the host of the free current), otherwise the material medium produces energy and delivers it to the EM fields. Similarly, the exchange of energy between the fields and the electric dipoles occurs at the rate of $\boldsymbol{E}\cdot\partial\boldsymbol{P}/\partial t$. Note that the bound current density $\partial\boldsymbol{P}/\partial t$ associated with the polarization $\boldsymbol{P}(\boldsymbol{r},t)$ behaves similarly to the free current density $\boldsymbol{J}_{\text{free}}$ with regard to any exchange of energy with the EM fields. This, however, is not the case with the magnetization $\boldsymbol{M}(\boldsymbol{r},t)$, whose exchange of energy with the fields is governed by the last term, $\boldsymbol{H}\cdot\partial\boldsymbol{M}/\partial t$, of Eq. (11). Clearly, it is *not* $\boldsymbol{J}_{\text{bound}}=\mu_{\text{o}}^{-1}\boldsymbol{\nabla}\times\boldsymbol{M}$ that participates in energy exchange with the $E$-field; rather, it is $\boldsymbol{J}_{\text{bound}}^{(\text{m})}=\partial\boldsymbol{M}/\partial t$ that is involved in the give and take of energy with the $H$-field. Here we see a first indication that the bound electric current density associated with $\boldsymbol{M}(\boldsymbol{r},t)$ behaves differently than the free current density $\boldsymbol{J}_{\text{free}}$ when it comes to interactions involving an exchange of energy. We will see further evidence of this deviant behavior when we discuss EM force and torque in Sec. 6. In the context of force and torque, we will also see evidence of a different behavior for $\boldsymbol{P}(\boldsymbol{r},t)$. Although the bound electric current density $\partial\boldsymbol{P}/\partial t$ associated with polarization continues to behave similarly to $\boldsymbol{J}_{\text{free}}$, we will find in Sec. 6 that the free and bound electric charge densities, $\rho_{\text{free}}$ and $-\boldsymbol{\nabla}\cdot\boldsymbol{P}$, no longer behave in similar ways.

**5. Linear and angular momentum densities of the electromagnetic field**. It is a well-known fact that EM fields carry momentum as well as energy. Perhaps the simplest way to deduce the magnitude of the field momentum from first principles is by means of the Einstein box thought experiment [4]. Shown in Fig. 1 is an empty box of length $L$ and mass $M$, placed on a frictionless rail, and free to move forward or backward. At some point in time, a blob of material attached to the left wall emits a short EM pulse of energy $\mathcal{E}$ and momentum $\boldsymbol{p}$, which remains collimated as it travels the length of the box and gets absorbed by another blob attached to the right-hand wall. The recoil velocity of the box is thus $-\boldsymbol{p}/M$, the time of flight is $L/c$, and the box displacement along the rail is $-(p/M)(L/c)$. Associating a mass $m=\mathcal{E}/c^2$ with the EM pulse and assuming that $M >> m$, it is easy to see that the displacement of the center-of-mass of the system is proportional to $(\mathcal{E}/c^2)L - M(p/M)(L/c)$. In the absence of external forces acting on the box, however, its center-of-mass is not expected to move. Setting the net displacement equal to zero, we find $p = \mathcal{E}/c$. Thus, in free space, a light pulse of energy $\mathcal{E}$ carries a momentum $p = \mathcal{E}/c$ along its direction of propagation. This result, which is independent of the particular shape of the pulse as a function of time, is accurate provided that the amplitude and phase profiles of the EM wave are fairly smooth and uniform over a large cross-sectional area, thus ensuring that the pulse remains collimated as it traverses the length of the box.

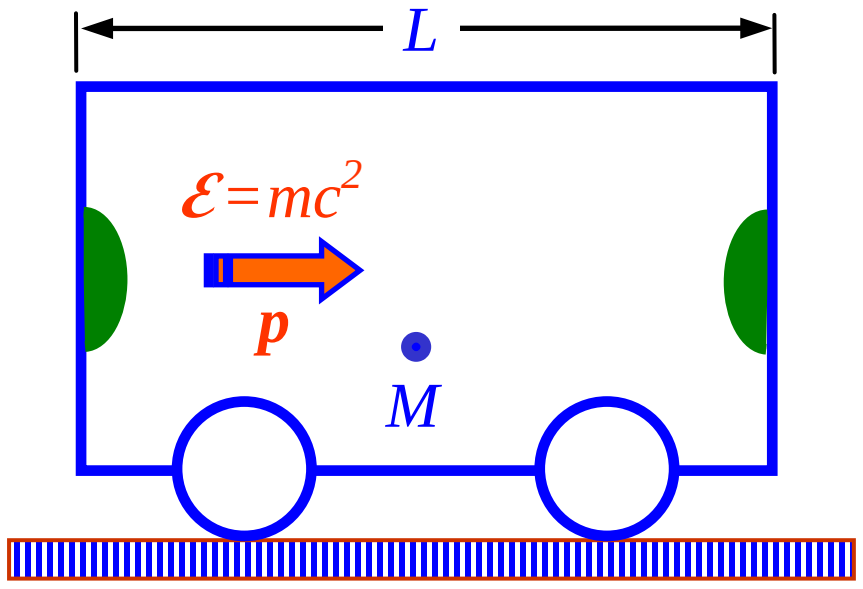

**Fig. 1**. Einstein box gedanken experiment.

In terms of the Poynting vector $\boldsymbol{S}$, one can readily show that the momentum density of the EM fields in the Einstein box thought experiment is given by $\boldsymbol{S}(\boldsymbol{r}, t)/c^2$. To see this, assume a cross-sectional area $A$ for the pulse, and note that the entire pulse moves at constant velocity $c$ from left to right. Choose an arbitrary cross-section of the pulse (perpendicular to its propagation direction), and observe that the EM energy passing through this cross-section during a short time interval $\Delta t$ is given by $\Delta\mathcal{E} = S(\boldsymbol{r}, t)A\Delta t$. This energy, which proceeds to occupy the infinitesimal volume $\Delta v = Ac\Delta t$ to the right of the chosen cross-section, yields an energy density $\Delta\mathcal{E}/\Delta v = S(\boldsymbol{r}, t)/c$ at point $\boldsymbol{r}$ at time $t$ and, consequently, a momentum density $\Delta\boldsymbol{p}/\Delta v = \boldsymbol{S}(\boldsymbol{r}, t)/c^2$ at that location.

The above arguments do not shed any light on the momentum of EM waves inside material media. However, a different thought experiment, due to N. L. Balazs and dating back to 1953 [7], reveals that the EM momentum density within a transparent slab must also be given by $\boldsymbol{S}(\boldsymbol{r}, t)/c^2$. Consider a transparent rod of length $L$, group refractive index $n_g$, and large mass $M$, as shown in Fig. 2. For purposes of the present argument, the rod may be made of an ordinary (dielectric) glass or a piece of transparent magnetic material. Let a short light pulse enter the rod from the left and exit from the right, without losses due to absorption or scattering, or due to reflections at the entrance and exit facets. Balazs suggested three different schemes for avoiding reflections at the facets [7], but, for our purposes, it suffices to assume the existence of perfect antireflection coatings on both facets.

When the pulse emerges from the rod, it will be delayed by the reduced speed of light within the material medium. In other words, had the pulse travelled parallel to its current path but outside the rod, it would have been ahead a distance of $(n_g - 1)L$ compared to where it will be upon emerging from the rod. Since there are no external forces acting on the system of rod plus light pulse, the center-of-mass of the system should be in the same location irrespective of whether the pulse went through the rod or followed a parallel path outside the rod.

Let the energy of the light pulse in vacuum be $\mathcal{E}$, corresponding to a mass of $\mathcal{E}/c^2$. The delay has caused a leftward shift of the product of mass and displacement by $(n_g - 1)L\mathcal{E}/c^2$. This must be compensated by a rightward shift of the rod itself. Let the light pulse have EM momentum $p$ while inside the rod. Considering that the momentum of the pulse before entering the rod is $\mathcal{E}/c$, the rod must have acquired a net momentum of $(\mathcal{E}/c)-p$ while the pulse travelled inside. Its net mass times forward displacement, therefore, must be $[(\mathcal{E}/c)-p]n_g L/c$. Equating the above rightward and leftward mass × displacements yields $p=\mathcal{E}/(n_g c)$ for the EM momentum of the pulse inside the rod. Note that the refractive index appearing in the above momentum expression is the group index (as opposed to the phase index) of the transparent medium. Once again, it is straightforward to show that the EM momentum density of the light pulse inside the slab must be given by $\boldsymbol{S}(\boldsymbol{r},t)/c^2$.

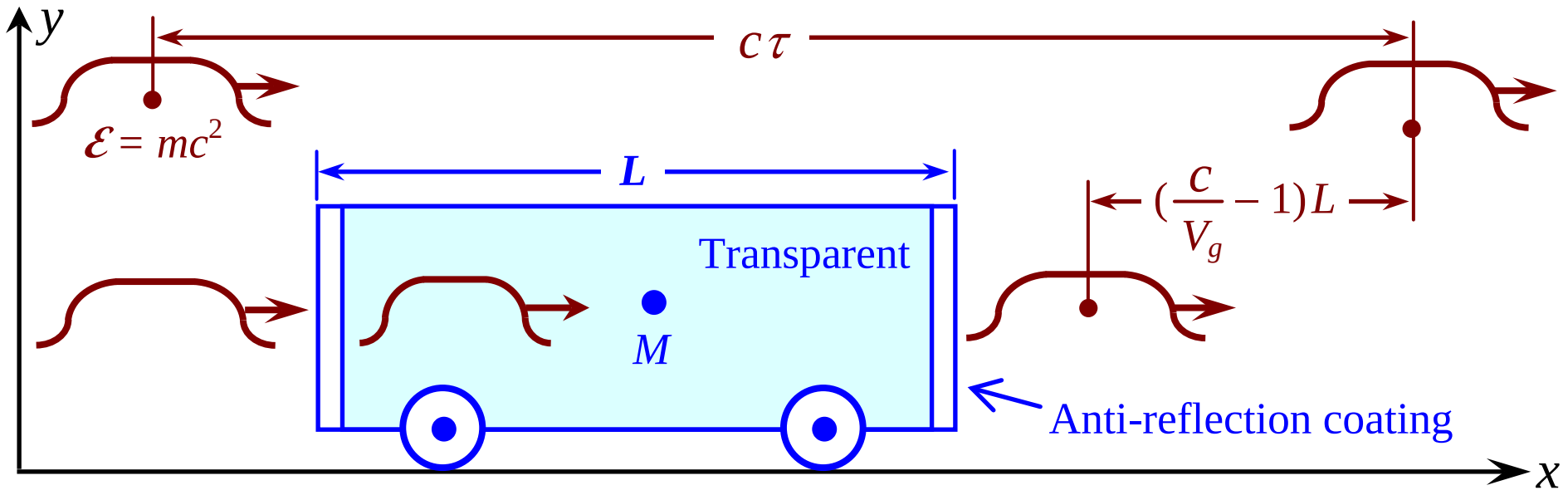

**Fig. 2**. The thought experiment of Balazs involves the propagation of a light pulse of energy $\mathcal{E}$ through a transparent rod of length $L$ and mass $M$. The rod can move on a frictionless rail along the $x$-axis. Since the group velocity $V_g=c/n_g$ of the pulse inside the rod is less than $c$, the emergent pulse is somewhat behind the location it would have reached had it travelled in vacuum all along.

A subtle point that requires careful consideration is that neither the Einstein box in Fig. 1 nor the transparent rod in Fig. 2 should be expected to behave as a perfectly rigid object. Since information cannot travel faster than the speed of light, any movement on the left-hand-side of an object cannot be communicated immediately to its right-hand-side. Therefore, one must analyze the motion of the Einstein box and also that of the Balazs rod not as the motion of a rigid block, but rather as a sequential movement of individual atomic layers within the block. We will not enter this discussion here as we have covered the technical issues at great length elsewhere [5,8]. Suffice it to say that, when the behavior of the material media is properly taken into consideration, the conclusions reached via the thought experiments of Einstein and Balazs remain intact.

The thought experiments of Einstein and Balazs indicate that the EM momentum density should be $\boldsymbol{S}(\boldsymbol{r},t)/c^2$, but the special circumstances of both experiments restrict the general validity of this conclusion. For instance, the light pulse under consideration must have a large cross-sectional diameter, so that it would not converge or diverge while propagating forward. Also, in the case of the Balazs experiment, since only the limited class of transparent materials has been considered, generalization to arbitrary media is not automatically justified. That is why we consider it a *postulate* of the classical theory that the EM momentum density under all circumstances is $\boldsymbol{p}(\boldsymbol{r},t)=\boldsymbol{S}(\boldsymbol{r},t)/c^2$. We shall refer to this important ansatz as the Abraham postulate [9]. By accepting the Abraham postulate, we reject all alternative expressions of the EM momentum density, including that due to H. Minkowski [10], for reasons to be elaborated in Sec. 7.

A corollary of the Abraham postulate is that the angular momentum density (with respect to the origin of coordinates) of EM fields under all circumstances is given by $\boldsymbol{L}(\boldsymbol{r},t)=\boldsymbol{r}\times\boldsymbol{S}(\boldsymbol{r},t)/c^2$. This angular momentum density pertains to both spin and orbital angular momenta of EM fields, as well as to any combination of these two types of angular momentum. In other words, whether the angular momentum is produced by circular polarization (spin), by vorticity (orbital), or by a combination of the two, the above expression for $\boldsymbol{L}(\boldsymbol{r},t)$ yields the total angular momentum density at any point $\boldsymbol{r}$ in space and any instant $t$ in time [11-23].

**6. Electromagnetic force and torque**. Just as the concept of energy for EM fields acquires significance when these fields interact with material media and, thereby, exchange energy with these media, the EM momentum of classical electrodynamics would also be meaningless unless there were interactions with material media in which linear and/or angular momentum could be exchanged. Generally speaking, such exchanges are mediated by force and torque exerted by the EM fields on material media. The force of $\boldsymbol{E}(\boldsymbol{r},t)$ and $\boldsymbol{B}(\boldsymbol{r},t)$ on a point-particle of charge $q$ and velocity $\boldsymbol{V}$ located at point $\boldsymbol{r}$ at instant $t$ is given by the Lorentz law of force [1,2], namely,

$$\boldsymbol{f}(\boldsymbol{r},t)=q[\boldsymbol{E}(\boldsymbol{r},t)+\boldsymbol{V}\times\boldsymbol{B}(\boldsymbol{r},t)]. \tag{12}$$

Application of the Lorentz law to an arbitrary distribution of charge and current densities yields the following expressions for the Lorentz force and torque densities:

$$\boldsymbol{F}_L(\boldsymbol{r},t)=\rho(\boldsymbol{r},t)\boldsymbol{E}(\boldsymbol{r},t)+\boldsymbol{J}(\boldsymbol{r},t)\times\boldsymbol{B}(\boldsymbol{r},t). \tag{13}$$

$$\boldsymbol{T}_L(\boldsymbol{r},t)=\boldsymbol{r}\times\boldsymbol{F}_L(\boldsymbol{r},t). \tag{14}$$

While the above equations work well with free-charge and free-current densities, it turns out that their (commonly practiced) application to bound charge and bound current densities associated with $\boldsymbol{P}$ and $\boldsymbol{M}$ creates a few problems. One of these problems is the need to introduce the concepts of "hidden energy" and "hidden momentum" into electrodynamics in order to make the theory compatible with energy and momentum conservation as well as with the special theory of relativity [24-36]. There exists, however, a modified version of Eq.(13), first proposed by A. Einstein and J. Laub in 1908 [37,38], which avoids the complications of hidden entities while remaining compatible with special relativity and with the conservation laws. The Einstein-Laub expression for EM force density, together with a corresponding expression for torque density, are written below:

$$\boldsymbol{F}_{EL}(\boldsymbol{r},t)=\rho_{\text{free}}\boldsymbol{E}+\boldsymbol{J}_{\text{free}}\times\mu_o\boldsymbol{H}+(\boldsymbol{P}\cdot\nabla)\boldsymbol{E}+(\partial\boldsymbol{P}/\partial t)\times\mu_o\boldsymbol{H}+(\boldsymbol{M}\cdot\nabla)\boldsymbol{H}-(\partial\boldsymbol{M}/\partial t)\times\varepsilon_o\boldsymbol{E}, \tag{15}$$

$$\boldsymbol{T}_{EL}(\boldsymbol{r},t)=\boldsymbol{r}\times\boldsymbol{F}_{EL}(\boldsymbol{r},t)+\boldsymbol{P}(\boldsymbol{r},t)\times\boldsymbol{E}(\boldsymbol{r},t)+\boldsymbol{M}(\boldsymbol{r},t)\times\boldsymbol{H}(\boldsymbol{r},t). \tag{16}$$

Equations (15) and (16) clearly show the action of the $\boldsymbol{E}$ and $\boldsymbol{H}$ fields on the four sources, $\rho_{\text{free}}$, $\boldsymbol{J}_{\text{free}}$, $\boldsymbol{P}$ and $\boldsymbol{M}$. Despite apparent differences between the Lorentz force and torque expressions of Eqs.(13,14) and the Einstein-Laub expressions in Eqs.(15,16), the two sets of equations are not all that far apart. One cosmetic difference is that the contribution of the momentum hidden in magnetic materials is present in the Lorentz formulation but absent from the Einstein-Laub formulation; see the Appendix for further detail. This is a distinction without a difference, however, since the EM momentum hidden in magnetic dipoles is not a physical observable, and, therefore, its removal from the Lorentz law of Eq.(13) does not constitute a change in the physics of the problem [8]. In fact, eliminating the hidden momentum associated with magnetic dipoles from the expressions of EM momentum density and the Lorentz force is similar to removing hidden energy from the expression of the Poynting vector, which is what has been done traditionally to keep the Poynting vector as $\boldsymbol{E}\times\boldsymbol{H}$ [1-3] rather than $\boldsymbol{E}\times\boldsymbol{B}/\mu_o$ [4]. With the contribution of the hidden momentum associated with magnetic dipoles removed from the Lorentz force and torque densities of Eqs.(13) and (14), it is shown in the appendix that the total force (and also total torque) exerted on any isolated object will turn out to be the same, whether one uses the Lorentz formulas or the Einstein-Laub expressions when computing the total force and torque.

The remaining differences between the two formulations have to do with the *distributions* of force and torque densities within material bodies; these differences can be observed only in deformable media. The choice between the Lorentz law (with hidden entities removed) and the Einstein-Laub law thus comes down to experimental observations of force and torque on deformable objects, and here the limited body of evidence appears to support the Einstein-Laub formulation. A good example is provided by the 1973 experiments of Ashkin and Dziedzic, where the surface deformations of a transparent liquid under a focused laser beam were measured [39]. The experimental results show good agreement with theoretical analyses based on the Einstein-Laub formula [40], but are in poor agreement with calculations based on the Lorentz law [41]. We mention in passing that, to date, the vast majority of theoretical analyses in the field of radiation pressure has been confined to transparent, non-magnetic dielectrics, and based on the following restricted version of the Einstein-Laub equation [42-52]:

$$\boldsymbol{F}_{EL}(\boldsymbol{r},t) = (\boldsymbol{P}\cdot\boldsymbol{\nabla})\boldsymbol{E} + (\partial\boldsymbol{P}/\partial t)\times\mu_0\boldsymbol{H}. \tag{17}$$

Discussions of hidden momentum and magnetic media aside, there remains a fundamental difference between the Lorentz and Einstein-Laub formulations of the EM force. While the Lorentz law treats the material media as continuous distributions of charge and current densities, the Einstein-Laub formulation tends to acknowledge the fact that, on atomic/molecular scales, all matter exhibits a degree of granularity that causes each atom/molecule to experience the force/torque of the local EM fields as individual electric/magnetic dipoles would. The schematic diagram in Fig. 3(a) makes it clear that, in a continuous model of a material medium, the force of the $E$-field on electric dipoles is exerted only on those regions where the bound charge density $-\boldsymbol{\nabla}\cdot\boldsymbol{P}$, is non-zero. In contrast, in the granular model depicted in Fig. 3(b), each atomic/molecular dipole experiences the force of the $E$-field in accordance with the force density expression $(\boldsymbol{P}\cdot\boldsymbol{\nabla})\boldsymbol{E}$. In general, the force density *distribution* in the continuum model differs substantially from that obtained for the discrete model, even though the *total* EM force exerted on any given object invariably turns out to be model-independent. We mention in passing that, in the discrete model of Fig. 3(b), the term $(\boldsymbol{P}\cdot\boldsymbol{\nabla})\boldsymbol{E}$ fails to capture the contribution of a locally uniform $E$-field to the torque exerted on a dipole, hence the need for the $\boldsymbol{P}\times\boldsymbol{E}$ term in the Einstein-Laub torque density equation (16). A similar argument holds for the $\boldsymbol{M}\times\boldsymbol{H}$ term.

With the caveat that the force and torque laws of Eqs. (15, 16) could benefit from further experimental verification, we believe the Einstein-Laub formulation provides universal expressions for the force and torque densities exerted by $\boldsymbol{E}$ and $\boldsymbol{H}$ on material media –media that can generally be described as spatio-temporal distributions of $\rho_{\text{free}}$, $\boldsymbol{J}_{\text{free}}$, $\boldsymbol{P}$ and $\boldsymbol{M}$ [53-55].

The force and torque densities of Eqs. (15) and (16), together with the linear and angular EM momentum densities described in the preceding section, are fully consistent with momentum conservation laws. For instance, in a closed system, the total Abraham momentum at a given instant of time may be evaluated as $\boldsymbol{p}_{EM}(t) = \iiint(\boldsymbol{E}\times\boldsymbol{H}/c^2)\,\mathrm{d}x\,\mathrm{d}y\,\mathrm{d}z$. Similarly, the total force exerted on the material media of the system may be obtained as $\boldsymbol{F}(t) = \iiint\boldsymbol{F}_{EL}(\boldsymbol{r},t)\,\mathrm{d}x\,\mathrm{d}y\,\mathrm{d}z$ using Eq. (15). It can then be shown under the most general conditions that $\boldsymbol{F}(t) = -\mathrm{d}\boldsymbol{p}_{EM}(t)/\mathrm{d}t$ [56]. A similar relation holds between the total torque exerted by EM fields on the material media and the total EM angular momentum of the system, namely, $\boldsymbol{T}(t) = -\mathrm{d}\boldsymbol{L}_{EM}(t)/\mathrm{d}t$.

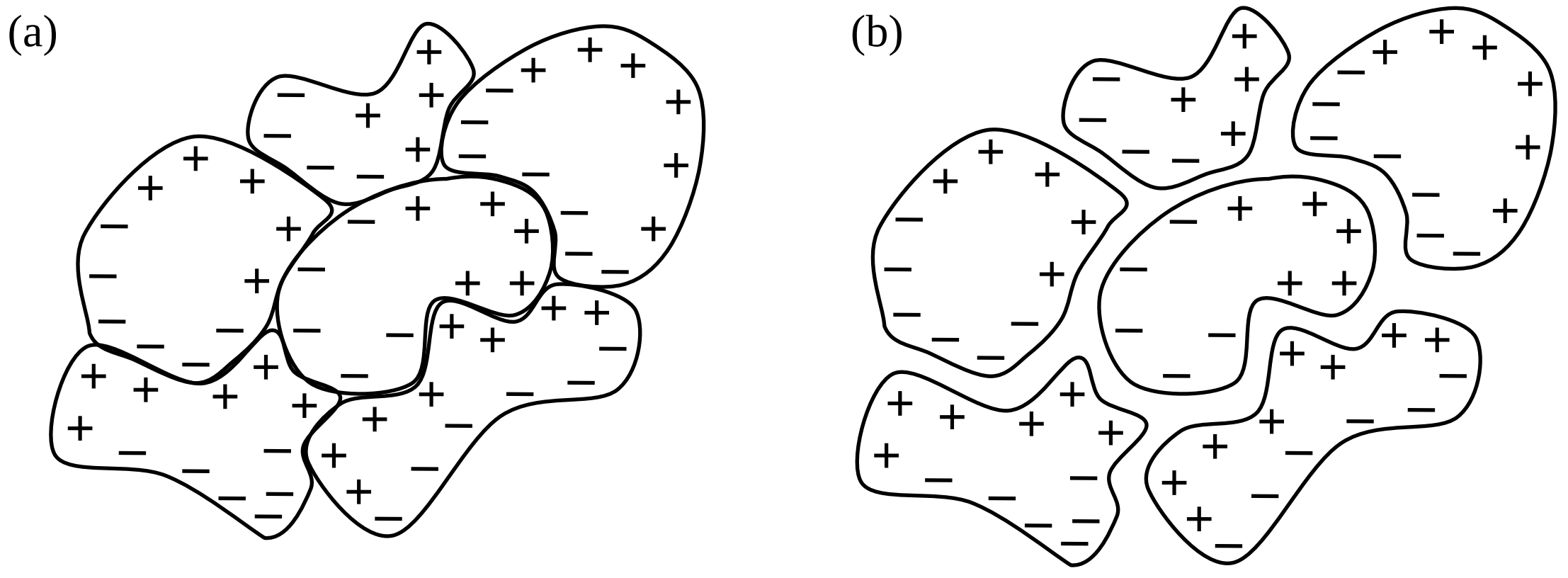

**Fig. 3**. (a) In a continuous medium, individual atoms/molecules are in contact with their adjacent neighbors. When these atoms/molecules are polarized, their positive and negative charges are largely cancelled out by the opposite charges associated with their nearest neighbors. The force of the $E$-field is thus exerted only on those regions where the net bound charge density, $-\boldsymbol{\nabla}\cdot\boldsymbol{P}$, is non-zero. (b) In a granular or discrete model of a material medium, a small gap separates each atom/molecule from its adjacent neighbors. The force of the $E$-field is thus experienced by individual dipoles, with a density given by $(\boldsymbol{P}\cdot\boldsymbol{\nabla})\boldsymbol{E}$. While the total EM force exerted on any given object is the same in the two formulations, the force density distributions, in general, are quite different.

**7. The Abraham-Minkowski controversy**. This controversy, which dates back to the first decade of the 20th century, has its roots in two formulations of the classical theory of electrodynamics. In M. Abraham's formulation, the momentum density is given by $\boldsymbol{E}\times\boldsymbol{H}/c^2$ [9], whereas H. Minkowski's approach requires the momentum density to be $\boldsymbol{D}\times\boldsymbol{B}$ [10]. In vacuum, the two formulations are equivalent and lead to a linear momentum $\boldsymbol{p}=(\mathcal{E}/c)\hat{\boldsymbol{z}}$ for a light pulse of energy $\mathcal{E}$ propagating along the $z$-axis, provided that the cross-sectional diameter of the beam is much greater than a wavelength. If the light pulse is also long enough that a single frequency $\omega$ could be assigned to it, one can say that the individual photons that comprise the pulse have energy $\hbar\omega$ and momentum $\hbar\omega/c$, where $\hbar$ is Planck's reduced constant.

When the above (long and wide) light pulse enters a transparent, non-dispersive dielectric medium of refractive index $n$, the Abraham momentum per photon becomes $\hbar\omega/(nc)$, whereas Minkowski's theory predicts the value of $n\hbar\omega/c$ for the photon momentum. In conjunction with the Balazs thought experiment [7], we argued in Sec. 5 that the Abraham momentum is the EM momentum of the light pulse inside the medium, and that the difference between the momentum of the pulse in vacuum and its EM momentum in a transparent medium is picked up by the medium as mechanical momentum. The presence of anti-reflection coatings on the medium depicted in Fig. 2, however, complicates the discussion – because of the forces exerted by the light pulse on these coatings [57] – so, for present purposes, we adopt a simpler slab without anti-reflection coatings, as shown in Fig. 4. Here the material of the slab is specified by its relative permittivity and permeability, $\varepsilon$ and $\mu$, both of which must be real-valued and have the same sign (i.e., both positive or both negative) for the slab to be transparent. The refractive index of such a material is given by $n=\sqrt{\mu\varepsilon}$, a positive real number for an ordinary material, but a negative real number for a negative-index material (i.e., one for which both $\mu$ and $\varepsilon$ are real and negative). The impedance $\sqrt{\mu/\varepsilon}$ of the medium, however, is real and positive in both cases.[‡] (The reader who is not already familiar with negative-index media may want to ignore all references to such media in the following discussion.)

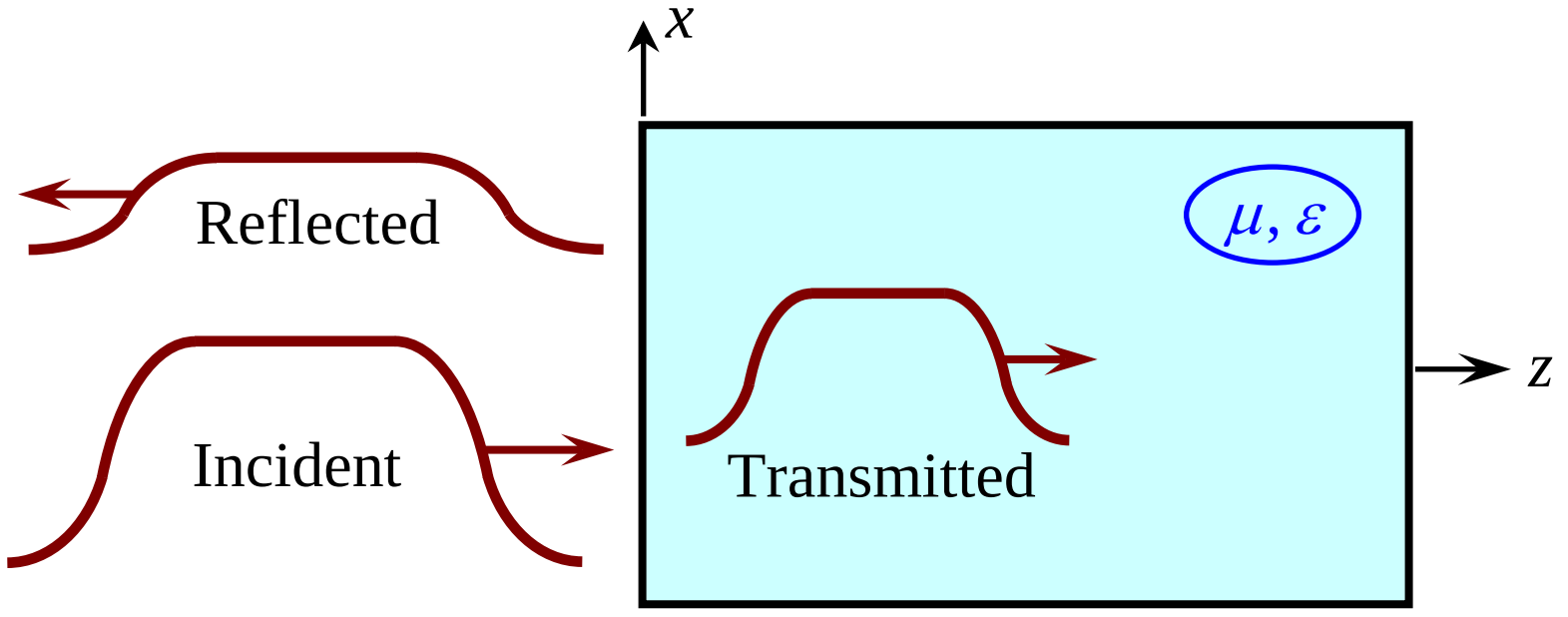

**Fig. 4**. A light pulse containing $N$ identical photons of energy $\hbar\omega$ is normally incident at the interface between free space and a magnetic slab of permittivity $\varepsilon$ and permeability $\mu$. The impedance of the medium is $\sqrt{\mu/\varepsilon}$, its refractive index $n=\sqrt{\mu\varepsilon}$, and its Fresnel reflection coefficient is $r$. Momentum conservation requires the transmitted photons to carry the momentum of the incident pulse *plus* that of the reflected pulse.

At normal incidence, the Fresnel reflection coefficient at the front facet of the slab is given by $r=(1-\sqrt{\varepsilon/\mu})/(1+\sqrt{\varepsilon/\mu})$. For an incident pulse of energy $\mathcal{E}$ and frequency $\omega$, the reflected and transmitted pulse energies will be $r^2\mathcal{E}$ and $(1-r^2)\mathcal{E}$, respectively. One may also think of $r^2$ as the fraction of incident photons that are reflected at the front facet (number of incident photons $N=\mathcal{E}/\hbar\omega$). Conservation of momentum requires the total momenta before and after incidence to be the same. Hence each photon inside the slab must carry a momentum of $(1+r^2)(\hbar\omega/c)/(1-r^2)$. This "photon momentum" now consists of two parts: an EM part, which is the Abraham momentum, and a mechanical part, which is associated with the motion of atoms constituting the slab. Expressing $r$ in terms of $\mu$ and $\varepsilon$ allows us to write the total (i.e., electromagnetic plus mechanical) momentum per photon inside the slab as follows:

$$p_{\text{photon}}=\tfrac{1}{2}\left(\sqrt{\varepsilon/\mu}+\sqrt{\mu/\varepsilon}\right)(\hbar\omega/c). \tag{18}$$

The above formula is valid for all real-valued $\varepsilon$ and $\mu$ of the same sign, representing transparent materials. In particular, for negative-index media the photon momentum is seen to be positive, i.e., directed along the positive $z$-axis. Note that the photon momentum in Eq.(18) is greater than or equal to the vacuum momentum $\hbar\omega/c$. Considering that the EM momentum per photon, following Abraham, is $\hbar\omega/(n_g c)$, where $n_g>1$ is the group refractive index of the host medium, the mechanical momentum of the photon will be the difference between its total momentum, given by Eq.(18), and its EM momentum.

Note that in the case of negative-index media, the refractive index $n=\sqrt{\mu\varepsilon}$ is a negative real number. However, this refractive index appears neither in Eq.(18), the expression of total photon momentum, nor in the expression of EM photon momentum, $\hbar\omega/(n_g c)$. Negative-index media are necessarily dispersive and, therefore, $n_g \neq n$ in such media. In contrast, a positive-index medium may or may not be dispersive, and, in the latter case, $n_g=n$. In a non-magnetic dielectric,

[‡] The choice of sign for the square roots involved here is not arbitrary. By allowing a small amount of absorption in the negative-index material, i.e., small imaginary components for $\varepsilon$ and/or $\mu$, one can see the rationale behind the choice of a negative sign for $n$. The positive sign for the impedance is justified when one considers the direction of flow of energy within the negative-index medium.

$\mu = 1$ and $n = \sqrt{\varepsilon}$, in which case the total photon momentum will be $p_{\text{photon}} = ½(n + n^{-1})(\hbar\omega/c)$, i.e., the arithmetic average of the Abraham and Minkowski momenta. If this material also happens to be non-dispersive, we will have $n_g = n > 1$; the EM momentum of the photon inside the slab will then be $\hbar\omega/(nc)$, resulting in a mechanical momentum of $½(n - n^{-1})(\hbar\omega/c)$. The bottom line is that the Minkowski momentum of a photon inside a transparent medium is neither its EM momentum, nor its mechanical momentum, nor its total momentum.

We emphasize here that all the results of the present section with regard to mechanical momentum can be derived directly from the Einstein-Laub force equation (15). In fact, everything about mechanical momentum, both linear and angular, that is exchanged between EM fields and material media can be calculated with the aid of Eqs. (15) and (16) [57-62].

It is our belief that the Abraham-Minkowski dilemma has been resolved in favor of the Abraham momentum [8, 62], although this opinion is not shared by everyone [63-68]. Several experimental observations are often cited in favor of the Minkowski momentum, among them measurements of radiation pressure on submerged mirrors, which apparently show the transfer of Minkowski momentum from photons in the host liquid to the submerged mirror [69-71]. We have shown elsewhere [72-74] that these observations can be readily explained on the basis of the classical theory of electrodynamics described in the preceding sections. In particular, we have argued that momentum transfer to the mirror is not the only exchange taking place in such experiments; some momentum is also transferred to the host liquid, and when everything is taken into account, the expression of total photon momentum in Eq. (18) is indisputably confirmed.

**8. Summary**. The material media of classical electrodynamics comprise spatio-temporal distributions of free charge and free current densities, $\rho_{\text{free}}(\boldsymbol{r}, t)$ and $\boldsymbol{J}_{\text{free}}(\boldsymbol{r}, t)$, as well as polarization $\boldsymbol{P}(\boldsymbol{r}, t)$ and magnetization $\boldsymbol{M}(\boldsymbol{r}, t)$. These constitute the four sources of the EM fields, which, in accordance with Maxwell's macroscopic equations, Eqs. (1)-(4), produce the fields $\boldsymbol{E}(\boldsymbol{r}, t)$ and $\boldsymbol{H}(\boldsymbol{r}, t)$. The displacement $\boldsymbol{D}(\boldsymbol{r}, t) = \varepsilon_0 \boldsymbol{E}(\boldsymbol{r}, t) + \boldsymbol{P}(\boldsymbol{r}, t)$ and the magnetic induction $\boldsymbol{B}(\boldsymbol{r}, t) = \mu_0 \boldsymbol{H}(\boldsymbol{r}, t) + \boldsymbol{M}(\boldsymbol{r}, t)$ are two additional fields that appear in Maxwell's equations.

The free charge and free current densities satisfy the continuity equation (5); moreover, $(\boldsymbol{J}_{\text{free}}, c\rho_{\text{free}})$ behaves as a 4-vector. Similar statements can be made about the bound charge and bound current densities, whether the bound entities are electric or magnetic in nature. The pairs $(\boldsymbol{P}, \boldsymbol{M})$, $(\boldsymbol{E}, \boldsymbol{B})$ and $(\boldsymbol{D}, \boldsymbol{H})$ form second rank tensors that can be readily Lorentz transformed between different inertial frames.

The only postulate of classical electrodynamics with regard to energy is the Poynting postulate, which states that the rate of flow of EM energy (per unit area per unit time) is given by the Poynting vector $\boldsymbol{S}(\boldsymbol{r}, t) = \boldsymbol{E}(\boldsymbol{r}, t) \times \boldsymbol{H}(\boldsymbol{r}, t)$. Subsequently, the Poynting theorem summarized in Eq. (11) provides expressions for the energy densities of the $\boldsymbol{E}$ and $\boldsymbol{H}$ fields, as well as the rates of exchange of energy between the fields and the material media.

The densities of the linear and angular momenta of EM fields are given by the Abraham postulate $\boldsymbol{p}(\boldsymbol{r}, t) = \boldsymbol{S}(\boldsymbol{r}, t)/c^2$ and its corollary $\boldsymbol{L}(\boldsymbol{r}, t) = \boldsymbol{r} \times \boldsymbol{S}(\boldsymbol{r}, t)/c^2$. The expression of EM angular momentum density pertains to both spin and orbital angular momenta, as well as to any combination of these two types of angular momentum.

Finally, the force and torque densities exerted by EM fields on material media are given by Eqs. (15) and (16), respectively. The Einstein-Laub force law of Eq. (15) and the corresponding torque law of Eq. (16) are consistent with the momentum conservation laws, in the sense that any change in the total EM linear/angular momentum of a closed system is precisely and instantaneously compensated by the mechanical linear/angular momentum imparted to the material media of the system via the exertion of force/torque over the entire system. There is no need to account for hidden entities (i.e., energy and momentum) in the above formulation of classical electrodynamics. The only caveat is that the long-neglected Einstein-Laub force density of Eq. (15) is still in need of rigorous experimental tests to confirm its differences with the Lorentz law of Eq. (13), differences that stem from the granularity of matter at atomic/molecular scales.

## Appendix

The density $\boldsymbol{F}_L(\boldsymbol{r},t)$ of the Lorentz force exerted on local charge and current densities, $\rho_{\text{total}}(\boldsymbol{r},t)$, $\boldsymbol{J}_{\text{total}}(\boldsymbol{r},t)$, is given by

$$\boldsymbol{F}_L(\boldsymbol{r},t) = \rho_{\text{total}}(\boldsymbol{r},t)\boldsymbol{E}(\boldsymbol{r},t) + \boldsymbol{J}_{\text{total}}(\boldsymbol{r},t)\times\boldsymbol{B}(\boldsymbol{r},t). \tag{A1}$$

Here $\rho_{\text{total}}(\boldsymbol{r},t) = \rho_{\text{free}}(\boldsymbol{r},t) - \boldsymbol{\nabla}\cdot\boldsymbol{P}(\boldsymbol{r},t)$ incorporates the densities of free and bound charges. Similarly, $\boldsymbol{J}_{\text{total}}(\boldsymbol{r},t) = \boldsymbol{J}_{\text{free}}(\boldsymbol{r},t) + \partial\boldsymbol{P}(\boldsymbol{r},t)/\partial t + \mu_o^{-1}\,\boldsymbol{\nabla}\times\boldsymbol{M}(\boldsymbol{r},t)$ contains the densities of free as well as bound currents arising from polarization and magnetization densities. The Lorentz force density of Eq. (A1) may thus be written

$$\boldsymbol{F}_L(\boldsymbol{r},t) = \rho_{\text{free}}\boldsymbol{E} - (\boldsymbol{\nabla}\cdot\boldsymbol{P})\boldsymbol{E} + \boldsymbol{J}_{\text{free}}\times\mu_o\boldsymbol{H} + (\partial\boldsymbol{P}/\partial t)\times\mu_o\boldsymbol{H} + (\boldsymbol{\nabla}\times\boldsymbol{M})\times\boldsymbol{H} + \boldsymbol{J}_{\text{total}}\times\boldsymbol{M}. \tag{A2}$$

The fact that Maxwell's Eq. (2) may be written equivalently as $\boldsymbol{\nabla}\times\boldsymbol{B} = \mu_o\boldsymbol{J}_{\text{total}} + \mu_o\varepsilon_o\partial\boldsymbol{E}/\partial t$, allows one to rewrite Eq. (A2) as follows:

$$\begin{aligned}\boldsymbol{F}_L = {}& \rho_{\text{free}}\boldsymbol{E} + \boldsymbol{J}_{\text{free}}\times\mu_o\boldsymbol{H} - (\boldsymbol{\nabla}\cdot\boldsymbol{P})\boldsymbol{E} + (\partial\boldsymbol{P}/\partial t)\times\mu_o\boldsymbol{H} + (\boldsymbol{\nabla}\times\boldsymbol{M})\times\boldsymbol{H} + (\boldsymbol{\nabla}\times\boldsymbol{H})\times\boldsymbol{M}\\ &+ \mu_o^{-1}(\boldsymbol{\nabla}\times\boldsymbol{M})\times\boldsymbol{M} - \varepsilon_o(\partial\boldsymbol{E}/\partial t)\times\boldsymbol{M}.\end{aligned} \tag{A3}$$

Then, with the help of the vector identity $\boldsymbol{\nabla}(\boldsymbol{A}\cdot\boldsymbol{B}) = (\boldsymbol{A}\cdot\boldsymbol{\nabla})\boldsymbol{B} + (\boldsymbol{B}\cdot\boldsymbol{\nabla})\boldsymbol{A} + \boldsymbol{A}\times(\boldsymbol{\nabla}\times\boldsymbol{B}) + \boldsymbol{B}\times(\boldsymbol{\nabla}\times\boldsymbol{A})$, we will have

$$\begin{aligned}\boldsymbol{F}_L(\boldsymbol{r},t) = {}& \rho_{\text{free}}\boldsymbol{E} + \boldsymbol{J}_{\text{free}}\times\mu_o\boldsymbol{H} - (\boldsymbol{\nabla}\cdot\boldsymbol{P})\boldsymbol{E} + (\partial\boldsymbol{P}/\partial t)\times\mu_o\boldsymbol{H} + (\boldsymbol{M}\cdot\boldsymbol{\nabla})\boldsymbol{H} + (\boldsymbol{H}\cdot\boldsymbol{\nabla})\boldsymbol{M} - \boldsymbol{\nabla}(\boldsymbol{M}\cdot\boldsymbol{H})\\ &+ \mu_o^{-1}(\boldsymbol{M}\cdot\boldsymbol{\nabla})\boldsymbol{M} - \tfrac{1}{2}\mu_o^{-1}\,\boldsymbol{\nabla}(\boldsymbol{M}\cdot\boldsymbol{M}) - (\partial\boldsymbol{M}/\partial t)\times\varepsilon_o\boldsymbol{E} - \varepsilon_o\partial(\boldsymbol{E}\times\boldsymbol{M})/\partial t.\end{aligned} \tag{A4}$$

Next, we combine and rearrange the terms and, noting from Maxwell's Eq. (4) that $\boldsymbol{\nabla}\cdot\boldsymbol{B} = 0$, we add the null term $\mu_o^{-1}(\boldsymbol{\nabla}\cdot\boldsymbol{B})\boldsymbol{M}$ to the right-hand side of Eq. (A4) to obtain

$$\begin{aligned}\boldsymbol{F}_L(\boldsymbol{r},t) = {}& \rho_{\text{free}}\boldsymbol{E} + \boldsymbol{J}_{\text{free}}\times\mu_o\boldsymbol{H} + (\boldsymbol{P}\cdot\boldsymbol{\nabla})\boldsymbol{E} + (\partial\boldsymbol{P}/\partial t)\times\mu_o\boldsymbol{H} + (\boldsymbol{M}\cdot\boldsymbol{\nabla})\boldsymbol{H} - (\partial\boldsymbol{M}/\partial t)\times\varepsilon_o\boldsymbol{E} - \varepsilon_o\partial(\boldsymbol{E}\times\boldsymbol{M})/\partial t\\ &- [(\boldsymbol{P}\cdot\boldsymbol{\nabla})\boldsymbol{E} + (\boldsymbol{\nabla}\cdot\boldsymbol{P})\boldsymbol{E}] + \mu_o^{-1}[(\boldsymbol{B}\cdot\boldsymbol{\nabla})\boldsymbol{M} + (\boldsymbol{\nabla}\cdot\boldsymbol{B})\boldsymbol{M}] - \mu_o^{-1}\,\boldsymbol{\nabla}[\boldsymbol{M}\cdot(\boldsymbol{B} - \tfrac{1}{2}\boldsymbol{M})].\end{aligned} \tag{A5}$$

The last three terms of the above equation are complete differentials, that is,

$$(\boldsymbol{P}\cdot\boldsymbol{\nabla})\boldsymbol{E} + (\boldsymbol{\nabla}\cdot\boldsymbol{P})\boldsymbol{E} = \partial(P_x\boldsymbol{E})/\partial x + \partial(P_y\boldsymbol{E})/\partial y + \partial(P_z\boldsymbol{E})/\partial z, \tag{A6a}$$

$$(\boldsymbol{B}\cdot\boldsymbol{\nabla})\boldsymbol{M} + (\boldsymbol{\nabla}\cdot\boldsymbol{B})\boldsymbol{M} = \partial(B_x\boldsymbol{M})/\partial x + \partial(B_y\boldsymbol{M})/\partial y + \partial(B_z\boldsymbol{M})/\partial z, \tag{A6b}$$

$$\boldsymbol{\nabla}[\boldsymbol{M}\cdot(\boldsymbol{B} - \tfrac{1}{2}\boldsymbol{M})] = [(\partial/\partial x)\hat{\boldsymbol{x}} + (\partial/\partial y)\hat{\boldsymbol{y}} + (\partial/\partial z)\hat{\boldsymbol{z}}][\boldsymbol{M}\cdot(\boldsymbol{B} - \tfrac{1}{2}\boldsymbol{M})]. \tag{A6c}$$

This means that, if $\boldsymbol{F}_L(\boldsymbol{r},t)$ of Eq. (A5) is integrated over the entire volume of space that contains material media, taking note of the fact that both $\boldsymbol{P}$ and $\boldsymbol{M}$ vanish outside the media, one finds that the net contribution to *total* force of the last three terms amounts to zero. (Their net contribution to total torque density with respect to the origin is given by $\boldsymbol{P}\times\boldsymbol{E} + \boldsymbol{M}\times\boldsymbol{H}$.) Thus, as far as the *total* force exerted on the material media of a given electromagnetic system is concerned, the last three terms of Eq. (A5) are inconsequential and may as well be dropped. We are then left with

$$\boldsymbol{F}_L(\boldsymbol{r},t) = \rho_{\text{free}}\boldsymbol{E} + \boldsymbol{J}_{\text{free}} \times \mu_o \boldsymbol{H} + (\boldsymbol{P}\cdot\boldsymbol{\nabla})\boldsymbol{E} + (\partial \boldsymbol{P}/\partial t)\times \mu_o \boldsymbol{H} + (\boldsymbol{M}\cdot\boldsymbol{\nabla})\boldsymbol{H} - (\partial \boldsymbol{M}/\partial t)\times \varepsilon_o \boldsymbol{E} - \varepsilon_o \partial(\boldsymbol{E}\times\boldsymbol{M})/\partial t. \tag{A7}$$

The last term in Eq. (A7) can be shown to be responsible for the so-called "hidden momentum." To ensure momentum conservation, it is thus necessary to remove from Eq. (A7) the term $-\varepsilon_o \partial(\boldsymbol{E}\times\boldsymbol{M})/\partial t$, which leaves behind the Einstein-Laub force density equation.

**Acknowledgements**. The author is grateful to Tobias S. Mansuripur, Ewan M. Wright, and Armis R. Zakharian for many helpful discussions.